\documentstyle [11pt]{article}
\newtheorem{Def}{Def.}[section]

\newtheorem{Lemma}[Def]{Lemma}

\newcommand{\Proof}{{\em{Proof: }}}
\newcommand{\QED}{\ \hfill $\FBox$ \\[1em]}

\newcommand{\M}{\mbox{${\cal M}$}}
\newcommand{\G}{\mbox{${\cal G}$}\ }
\newcommand{\Lsa}{\mbox{$L^{\mbox{\scriptsize sa}} $}}
\newcommand{\Tr}{\mbox{Tr\/}}
\newcommand{\tr}{\mbox{tr\/}}
\newcommand{\Pd}[1]{ \frac{\partial}{\partial x^{#1}} }
\newcommand{\Pdn}[1]{ \frac{\partial}{\partial #1} }
\newcommand{\Pdd}{\mbox{$\partial$ \hspace{-1.2 em} $/$}}
\newcommand{\Op}{{\cal O}\!{\cal P}}
\newcommand{\X}{X}
\newcommand{\Sl}{\mbox{$\prec \!\!$ \nolinebreak}}
\newcommand{\Sr}{\mbox{\nolinebreak $\succ$}}
\newcommand{\bra}{\mbox{$< \!\!$ \nolinebreak}}
\newcommand{\ket}{\mbox{\nolinebreak $>$}}
\newcommand{\kernel}{\mbox{kernel}}
\newcommand{\spc}{\;\;\;\;\;\;\;\;\;\;}
\newcommand{\T}{{\mbox{T }}}
\newcommand{\texp}{{\mbox{Texp }}}
\newcommand{\arctanh}{{\mbox{arctanh}}}
\newcommand{\const}{{\mbox{const }}}

\newcommand{\C}{\mbox{\rm I \hspace{-1.25 em} {\bf C}}}
\newcommand{\R}{\mbox{\rm I \hspace{-.8 em} R}}
\newcommand{\1}{\mbox{\rm 1 \hspace{-1.05 em} 1}}
\newcommand{\Z}{\mbox{\rm \bf Z}}
\newcommand{\sR}{\mbox{\rm \scriptsize I \hspace{-.8 em} R}}
\newcommand{\N}{\mbox{\rm I \hspace{-.8 em} N}}
\newcommand{\sN}{\mbox{\rm \scriptsize I \hspace{-.8 em} N}}
\newcommand{\sZ}{\mbox{\rm \scriptsize \bf Z}}
\newcommand{\loc}{\mbox{\rm{\scriptsize{loc}}}}
\newcommand{\nab}{\breve{\nabla}}
\newcommand{\rhe}{\rho_{\mbox{\rm \scriptsize em}}}
\newcommand{\rhm}{\rho_{\mbox{\rm \scriptsize m}}}
\newcommand{\sj}{\breve{\jmath}}
\newcommand{\srhe}{\breve{\rho}_{\mbox{\rm \scriptsize em}}}
\newcommand{\srhm}{\breve{\rho}_{\mbox{\rm \scriptsize m}}}

\newcommand{\Equ}[1]{\begin{equation} \label{eq:#1}}
\newcommand{\EndEqu}{\end{equation}}
\newcommand{\Ref}[1]{(\ref{eq:#1})}

\newcommand{\FBox}{\rule{2mm}{2.25mm}}
\newcommand{\OBox}{\raisebox{.6ex}{\fbox{}}\,}
\newcommand{\A}{{\mbox{${\cal A}$}}}
\newcommand{\B}{{\mbox{${\cal B}$}}}

\title{Ground State Structure of a Coupled 2-Fermion System in
Supersymmetric Quantum Mechanics}
\author{Felix Finster\\ Harvard University, Department of Mathematics}
\date{November 1996}

\begin{document}
\maketitle

\begin{abstract}
We prove the uniqueness of the ground state for a supersymmetric
quantum mechanical system of two fermions and two bosons, which is
closely related to the $N=1$ WZ-model.
The proof is constructive and gives detailed information on what
the ground state looks like.
\end{abstract}

\section{Introduction, definition of the model}
A supersymmetric quantum system is defined by two self-adjoint
operators $Q, \gamma$ on a Hilbert space ${\cal{H}}$ satisfying
the relations
\[ \gamma^2\:=\: 1 \;\;\;\;\;,\spc \left\{\gamma,\: Q \right\} \;=\; 0
	\spc . \]
They are called supercharge and grading operator.
The Hamiltonian is given by $H:=Q^2 \geq 0$.
It is convenient to decompose ${\cal{H}}={\cal{H}}^+ \oplus
{\cal{H}}^-$ into the eigenspaces of $\gamma$, which leads to the
$(2 \times 2)$-matrix representations
\[ \gamma \;=\; \left( \begin{array}{cc} 1 & 0 \\ 0 & -1
	\end{array} \right) \;\;,\;\;\;\;\;
Q \;=\; \left( \begin{array}{cc} 0 & Q^- \\ Q^+ & 0
	\end{array} \right) \;\;,\;\;\;\;\;
H \;=\; \left( \begin{array}{cc} Q^- \: Q^+ & 0 \\ 0 & Q^+ \: Q^-
	\end{array} \right) \spc . \]
Eigenstates of $H$ with zero eigenvalue are called ground states.
The ground state structure of supersymmetric systems has been studied
intensively (see e.g. 
\cite{CH},\cite{JW},\cite{JLW1},\cite{JLW2},\cite{JLW3}).
It is of particular interest because it tells about a spontaneous
breaking of the supersymmetry (see e.g. \cite{C}).
In this paper we will analyze the system where $Q^\pm$ are the differential
operators on $L^2(\R^2)^2$
\begin{eqnarray}
\label{eq:0a}
Q^+ &=& -i \left( \begin{array}{cc} \partial_z & 0 \\ 0 & \partial_{\bar{z}}
	\end{array} \right) \:-\: i \left( \begin{array}{cc} 0 & 
\partial_z W
	\\ \partial_{\bar{z}} \overline{W} & 0 \end{array} \right) \\
\label{eq:0b}
Q^- &=& (Q^+)^* \;=\;
	-i \left( \begin{array}{cc} \partial_{\bar{z}} & 0 \\ 0 & \partial_z
	\end{array} \right) \:+\: i \left( \begin{array}{cc} 0 & 
\partial_z W
	\\ \partial_{\bar{z}} \overline{W} & 0 \end{array} \right)
\end{eqnarray}
(we use the complex notation
$z=x+iy$, $\bar{z}=x-iy$, $\partial_z = \frac{1}{2}
(\partial_x - i \partial_y)$, $\partial_{\bar{z}} = \frac{1}{2}
(\partial_x + i \partial_y)$).\\
The superpotential $W=W(|z|)$ is assumed to be a complex polynomial of
degree at least two.

As motivation we explain, in which context this system arises
and why it is of some general interest:
One of the simplest examples of a supersymmetric quantum field theory
is the $N=1$ WZ-model on the cylinder. It is defined by
\[ Q \;=\; \int_{S^1} dx \left(
	\Psi_1(x) \: \pi(x) - \Psi_2(x) \: \partial_x \varphi(x) \:+\:
	\Psi_2(x) \: V^\prime(\varphi(x)) \right) \spc , \]
where the field operators $\Psi_j, \pi, \varphi$
satisfy the canonical (anti-)commutation relations
\begin{eqnarray*}
\left\{ \Psi_i(x), \:\Psi_j(y) \right\} &=& \delta_{ij} \:
	\delta(x,y) \;\;,\;\;\;\;\;
   \left[\pi(x),\:\varphi(y) \right] \;=\; -i \: \delta(x,y) \\
\left[\Psi_i(x), \pi(y) \right] &=& \left[ \Psi_i(x),\: \varphi(y)
	\right] \;=\; 0 \spc .
\end{eqnarray*}
The analysis of the ground state structure of this system turns out to be an
involved problem. If the potential $V$ is sufficiently small, cluster
expansion methods can be used to prove the uniqueness of the ground
state in the infinite volume limit (see \cite{W}).
The general difficulty is the complicated coupling of the quantum
fields. To make this more apparent, we expand the field operators in a
Fourier series
\[ f(x) \;=\; \frac{1}{\sqrt{2 \pi}} \sum_{k \in \sZ} f(k) \: e^{-ikx}
	\spc , \]
which leads to the equation
\Equ{1}
Q \;=\; \sum_{k \in \sZ} \left(
	\Psi_1(k) \: \pi(-k) \:-\: ik \: \Psi_2(k) \: \varphi(-k)
	\:+\: \Psi_2(k) \: V^\prime(-k) \right) \spc .
\EndEqu
The Fourier modes are mixed up in the sum and
cannot be easily separated. We can view the operators $\Psi_1(k), \Psi_2(-k)$
as the annihilation/creation-operators of a fermion with momentum $k$,
conversely $\pi(k), \varphi(-k)$ are the canonical variables of a
$k$-momentum boson. With this picture, \Ref{1} describes a strongly 
interacting
fermion/boson-system.

A rough simplification of the problem consists in the reduction to the
zero modes,
\[ Q \;=\; \Psi_1(0) \: \pi(0) \:+\: \Psi_2(0) \: V^\prime(0) \spc . \]
This is $N=1$ quantum mechanics
\cite{JLL}, as one sees after rewriting $\varphi, \pi$ as canonical
multiplication/differential operators on $L^2(\R)$ and choosing a
matrix representation for the fermionic operators $\Psi_j$.
Unfortunately the coupling structure of the different Fourier modes in
\Ref{1} gets completely lost in this limit.
Therefore it seems interesting to study the system of two coupled
Fourier modes,
\[ Q \;=\; \sum_{k = \pm1} \left(
	\Psi_1(k) \: \pi(-k) \:-\: ik \: \Psi_2(k) \: \varphi(-k)
	\:+\: \Psi_2(k) \: V^\prime(-k) \right) \spc . \]
This model shows many characteristics of the field theory
\Ref{1}, but is still simple enough for a detailed analysis.
In a suitable representation of the field operators, it coincides
with the system \Ref{0a}, \Ref{0b} if we choose $W(r)=r^2 + i V(r)$.

We conclude that the differential operators \Ref{0a}, \Ref{0b}
describe a supersymmetric quantum mechanical system of two fermions
and two bosons. The coupling is typical for supersymmetric field
theories and should be helpful for
the understanding of these systems.

\section{Reduction to a one-dimensional problem}
We start the analysis of the ground states with some explicit
calculations.
Since the equations $H \Psi = 0$, $Q \Psi = 0$ are equivalent, it
suffices to look for solutions of the equations
\[ Q^\pm \: \Psi^\pm \;=\; 0 \spc{\mbox{with}}\spc
	\Psi^\pm \in L^2(\R^2)^2 \spc . \]
Because of the radial symmetry of $W$, it is useful to choose polar 
coordinates
$(r, \varphi)$. We have
\[ \partial_z \;=\; \frac{e^{-i \varphi}}{2} \: \left(
	\partial_r - \frac{i}{r} \:\partial_\varphi \right) \;\;\;,\spc
\partial_{\bar{z}} \;=\; \frac{e^{i \varphi}}{2} \: \left(
	\partial_r + \frac{i}{r} \:\partial_\varphi \right) \]
and thus
\begin{eqnarray*}
Q^+ &=& -\frac{i}{2} \left( \begin{array}{cc}
	e^{-i \varphi} \: \partial_r & e^{-i \varphi} \: W^\prime \\
	e^{i \varphi} \: \overline{W}^\prime & e^{i \varphi} \: \partial_r
	\end{array} \right) \:+\: \frac{1}{2r} \left(
	\begin{array}{cc} -e^{-i \varphi} & 0 \\
	0 & e^{i \varphi} \end{array} \right) \partial_\varphi\\
Q^- &=& -\frac{i}{2} \left( \begin{array}{cc}
	e^{i \varphi} \: \partial_r & -e^{-i \varphi} \: W^\prime \\
	-e^{i \varphi} \: \overline{W}^\prime & e^{-i \varphi} \: \partial_r
	\end{array} \right) \:+\: \frac{1}{2r} \left(
	\begin{array}{cc} e^{i \varphi} & 0 \\
	0 & -e^{-i \varphi} \end{array} \right) \partial_\varphi \spc .
\end{eqnarray*}
We multiply the equation $Q^+ \:\Psi^+=0$ with the matrix
$2i \:{\mbox{diag}}(e^{i \varphi}, e^{-i \varphi})$,
\[ \left[ \partial_r \:-\: \frac{i}{r} \left( \begin{array}{cc}
	1 & 0 \\ 0 & -1 \end{array} \right) \:\partial_\varphi \:+\:
	\left( \begin{array}{cc} 0 & W^\prime \\ \overline{W}^\prime & 0
	\end{array} \right) \right] \Psi^+ \;=\; 0 \spc . \]
The differential operator in the square bracket commutes with
angular momentum $i \partial_\varphi$. Thus we can make the separation
ansatz
\Equ{2}
\Psi^+(r, \varphi) \;=\; \frac{1}{\sqrt{2 \pi}} \:
	\sum_{m \in \sZ} \Psi^+_m(r) \: e^{-im \varphi}
\EndEqu
and obtain the ordinary differential equations $A_m \Psi^+_m = 0$ with
\Equ{8}
A_m(r) \;=\; \frac{d}{dr} \:-\: \frac{m}{r} \left( \begin{array}{cc} 1 & 
0 \\
	0 & -1 \end{array} \right) \:+\: \left( \begin{array}{cc}
	0 & W^\prime \\ \overline{W}^\prime & 0 \end{array} \right) \spc .
\EndEqu
For the equation $Q^- \:\Psi^- =0$ we substitute
\Equ{3}
\Psi^-(r,\varphi) \;=\; \frac{1}{r} \left( \begin{array}{cc}
	e^{-i \varphi} & 0 \\ 0 & -e^{i \varphi} \end{array} \right) \:
	\Phi^-(r, \varphi) \spc .
\EndEqu
The resulting equation for $\Phi^-$ has, after multiplication with
the matrix $2i r\:{\mbox{diag}} (1,-1)$, the form
\[ \left[ \partial_r \:+\: \frac{i}{r} \left( \begin{array}{cc}
	1 & 0 \\ 0 & -1 \end{array} \right) \:\partial_\varphi \:+\:
	\left( \begin{array}{cc} 0 & W^\prime \\ \overline{W}^\prime & 0
	\end{array} \right) \right] \Phi^- \;=\; 0 \spc . \]
After separating the variables by
\Equ{4}
\Phi^-(r, \varphi) \;=\; \frac{1}{\sqrt{2 \pi}} \:
	\sum_{m \in \sZ} \Phi^-_m(r) \: e^{im \varphi} \spc ,
\EndEqu
the radial functions $\Phi^-_m$ must satisfy the equations $A_m \Phi^-_m=0$.

In this way we have reduced the problem to the analysis of the ordinary
differential equations
\Equ{5}
A_m \: f_m \;=\; 0 \;\;\;\;,\spc f \in C^1(\R^+)^2 \spc .
\EndEqu
The conditions $\Psi^\pm \in L^2(\R^2)^2$ mean, according to
\Ref{2}, \Ref{3}, \Ref{4}, that the solutions $\Psi^+_m$, $\Phi^-_m$ must
satisfy the conditions
\begin{eqnarray}
\label{eq:6}
\int_0^\infty |\Psi^+_m|^2 \; r \: dr &\leq& \infty \\
\label{eq:7}
\int_0^\infty |\Phi^-_m|^2 \; \frac{dr}{r} &\leq& \infty \spc .
\end{eqnarray}
It remains to find out for which values $m \in \Z$ there are solutions
of \Ref{5} satisfying \Ref{6} resp.\ \Ref{7}.

\section{A differential inequality independent of $W$}
We define for a given solution $f_m$ of \Ref{5} the real function
$\Lambda_{f_m}$ by
\[ \Lambda_{f_m}(r) \;=\; \bra f_m(r),\:
	\left( \begin{array}{cc} 1 & 0 \\ 0 & -1 \end{array} \right)
	 f_m(r) \ket \spc . \]
It satisfies the equation
\begin{eqnarray}
\frac{d}{dr} \Lambda_{f_m} &=& \frac{2m}{r} \: \bra f_m, \:f_m \ket
	\nonumber \\
&&-\: \bra \left( \begin{array}{cc} 0 & W^\prime \\
	\overline{W}^\prime & 0 \end{array} \right) f_m, \:
	\left( \begin{array}{cc} 1 & 0 \\ 0 & -1 \end{array} \right) 
\:f_m \ket
	\:-\: \bra f_m,\: \left( \begin{array}{cc} 1 & 0 \\
	0 & -1 \end{array} \right) \left( \begin{array}{cc} 0 & W^\prime \\
	\overline{W}^\prime & 0 \end{array} \right) f_m \ket \nonumber \\
\label{eq:8a}
&=& \frac{2m}{r} \: \bra f_m, \:f_m \ket
\end{eqnarray}
and thus
\Equ{12}
\frac{d}{dr} \Lambda_{f_m} \;\geq\; \frac{2m}{r} \: |\Lambda_{f_m}| \spc .
\EndEqu
This inequality implies that $m$ vanishes for all ground states:
\begin{Lemma}
For $m \neq 0$ all (nontrivial) solutions of \Ref{5} violate both \Ref{6}
and \Ref{7}.
\end{Lemma}
{\Proof}
Suppose that $m \neq 0$ and that a solution $f_m$ of \Ref{5} satisfies
either \Ref{6} or \Ref{7}.
We can assume that $m>0$, because we can otherwise replace $\Lambda_{f_m}$
by $-\Lambda_{f_m}$ and multiply \Ref{8a} by a factor $-1$.
According to \Ref{12}, the function $\Lambda_{f_m}$ is monotonely
increasing.

Suppose that there is $r_0>0$ with $\Lambda_{f_m}(r_0) >0$. Then
$\Lambda_{f_m}(r) > 0$ for all $r>r_0$ and
\[ \frac{d}{dr} \Lambda_{f_m}(r) \;\geq\; \frac{2m}{r} \: \Lambda_{f_m}(r)
	\spc . \]
Integrating this inequality yields the bound
\[ \Lambda_{f_m}(r) \;\geq\; \Lambda_{f_m}(r_0) \: \left(
	\frac{r}{r_0} \right)^{2m} \spc {\mbox{for $r \geq r_0$}} . \]
Since
$|f_m(r)|^2 \geq \Lambda_{f_m}(r)$, the function $|f_m(r)|^2$ grows at
least quadratic at infinity, which contradicts \Ref{6}, \Ref{7}.

Suppose conversely that there is $r_0>0$ with $\Lambda_{f_m}(r_0) < 0$.
Then $\Lambda_{f_m}(r)<0$ for all $r < r_0$ and
\[ \frac{d}{dr} \Lambda_{f_m}(r) \;\geq\; -\frac{2m}{r} \: \Lambda_{f_m}(r)
	\spc . \]
We thus have the bound
\[ \Lambda_{f_m}(r) \;\leq\; \Lambda_{f_m}(r_0) \: \left(
	\frac{r_0}{r} \right)^{2m} \spc , \]
which imples that $|f_m(r)|^2$ has at least a quadratic pole
at the origin. This is again a contradiction to \Ref{6}, \Ref{7}.

We conclude that $\Lambda_{f_m}$ must vanish identically. But then,
according to \Ref{8a}, $f_m \equiv 0$.
\QED
According to this Lemma it remains to analyze the equation $A_0 \:f=0$.
Since the operator $A_0$ is regular for $r=0$, the solutions $f$ have
no singularities at the origin, $f \in C^1(\R^+_0)$.

\section{The equation $Q^- \Psi^- = 0$ only has the trivial solution}
Let $\Phi^-_0$ be a solution of the equation $A_0 \:\Phi^-_0=0$ satisfying
\Ref{7}. Because of the divergent factor $r^{-1}$ in \Ref{7},
$\Phi^-$ must vanish at the origin.
Hence, by the uniqueness theorem for ordinary differential equations,
$\Phi^-_0 \equiv 0$.

\section{The equation $Q^+ \Psi^+ = 0$ has a unique solution}
Let $\Psi^+_0$ be a solution of the equation $A_0 \:\Psi^+_0=0$ satisfying
\Ref{6}.

We use the scalar equation \Ref{8a} to reduce the degrees of freedom
of $\Psi^+_0$:
For $m=0$, \Ref{8a} implies that $\Lambda_{\Psi^+_0}(r)$ is a constant.
Since $|\Psi^+_0|^2| \geq \Lambda_{\Psi^+_0}$, \Ref{6} can only be satisfied
if this constant is zero, $\Lambda_{\Psi^+_0} \equiv 0$.
Hence we can represent $\Psi^+_0$ in the form
\Equ{17}
\Psi^+_0(r) \;=\; \left( \begin{array}{c} u(r) \\ u(r) \: e^{-i \alpha(r)}
	\end{array} \right)
\EndEqu
with a complex function $u(r)$ and a real function $\alpha(r)$.
Furthermore we choose a polar representation
\[ W^\prime = U(r) \: e^{i \beta(r)} \]
of $W^\prime$ with smooth real functions $U, \beta$. We can assume
that $U(r)$ is positive for large $r$.
After these transformations, the equation $A_0 \Psi^+_0$ takes the form
\begin{eqnarray*}
\frac{d}{dr} \ln(u) &=& -U \: e^{-i(\alpha - \beta)} \\
\frac{d}{dr} \ln(u) \:-\: i \alpha^\prime &=& -U \: e^{i (\alpha - \beta)}
	\spc .
\end{eqnarray*}
It will turn out to be more convenient to choose $\gamma:=\alpha-\beta$
as the variable describing the relative phase in \Ref{17}.
Then we get the system of differential operators
\begin{eqnarray}
\label{eq:18}
\gamma^\prime &=& 2 U \:\sin(\gamma) \:-\: \beta^\prime \\ 
\label{eq:19}
\frac{d}{dr} \ln u &=& -U \: e^{-i \gamma} \spc .
\end{eqnarray}
The condition \Ref{6} transforms into
\Equ{20}
\int_0^\infty |u|^2 \: r \: dr \;<\; \infty \spc .
\EndEqu

This construction has simplified the equations considerably.
Since \Ref{18} does not depend on $u$, we can study the equation
for $\gamma$ independently.
For given $\gamma$, we can solve \Ref{19} by integration,
\Equ{17a}
u(r) \;=\; c \: \exp \left( -\int_0^r U(\tau) \: e^{-i \gamma(\tau)} \:
	d\tau \right) \;\;\;\;\;,\spc c \in \C .
\EndEqu
The condition \Ref{20} can be reformulated with the asymptotic
behaviour of $\gamma$:
Since \Ref{18},\Ref{19} are regular at the origin, \Ref{20} only yields
a condition on the decay of $u$ at infinity.
Suppose that $\gamma(r) \in (-\frac{\pi}{2}, \frac{\pi}{2}) \;({\mbox{mod }}
2 \pi)$ for large $r$. Then \Ref{17a} decays exponentially at infinity,
and \Ref{20} will be satisfied.
If, conversely, $\gamma(r) \in (\frac{\pi}{2}, \frac{3 \pi}{2}) \;
({\mbox{mod }} 2 \pi)$ for large $r$, the function $u(r)$ will grow
exponentially, and \Ref{20} will be violated.

We have thus reduced the problem to the analysis of the nonlinear
scalar equation \Ref{18}.
It is useful to view this equation as a dynamical system.
In the special case $\beta^\prime \equiv 0$, the equation
$\gamma^\prime = 2 U \: \sin(\gamma)$ has the two fixed points
$\gamma =0, \pi\;({\mbox{mod }}
2 \pi)$, which are repulsive and
attractive respectively. Every solution $\gamma$ with $\gamma(r) \neq 0$
will run into the attractive fixed point, $\lim_{r \rightarrow \infty}
\gamma(r) = \pi$.
We conclude that \Ref{20} implies that $\gamma \equiv 0\;({\mbox{mod }}
2 \pi)$.
The equation for $u$ then has (up to normalization) the unique
solution
\[ u(r) \;=\; \exp \left( -\int_0^r U(\tau) \: d\tau \right) \spc , \]
and we thus have exactly one ground state.

In the general case, equation \Ref{18} is more complicated.
Fortunately we can apply a perturbation argument:
In the limit $r \rightarrow \infty$, $\beta(r)$ converges to a phase
which is determined by the coefficient of the highest power of $W$.
As consequence, $\beta^\prime$ decays at least quadratic at infinity, i.e.
\Equ{21}
|\beta^\prime(r)| \;\leq\; \frac{C}{r^2}
\EndEqu
for a suitable constant $C$.
Since we must only consider the situation for large $r$, we can
view the summand $-\beta^\prime$ in \Ref{18}
as being arbitrarily small. Therefore we can still expect
a unique solution $\gamma$ with $\lim_{r \rightarrow \infty} \gamma(r) = 0
\;({\mbox{mod }} 2 \pi)$.
This consideration is made mathematically precise in the following
lemma.
\begin{Lemma}
The equation \Ref{18} has a unique solution $\gamma$ with
$\lim_{r \rightarrow \infty} \gamma(r) =0 \;({\mbox{mod }} 2 \pi)$.
For all other solutions $\bar{\gamma}$, $\lim_{r \rightarrow \infty}
\bar{\gamma}(r) = \pi \;({\mbox{mod }} 2 \pi)$.
\end{Lemma}
{\Proof}
We define the (nonlinear) operator $N$ by
\[ N \;:\; H^{1,1}(\R^+) \rightarrow L^1(\R^+) \;:\;
	f \rightarrow f^\prime - \sin (f) \spc . \]
$N$ is differentiable at the origin and $dN_{|0} f = f^\prime - f$.
For $g \in L^1(\R^+)$ let $f$ be the function
\[ f(y) \;=\; - \int_y^\infty dx \; e^{y-x} \: g(x) \spc . \]
The calculation
\begin{eqnarray*}
\int_0^\infty |f(y)| \: dy &\leq& \int_0^\infty dy \int_y^\infty dx \;
	e^{y-x} \: |g(x)|
\;=\; \int_0^\infty dx \int_0^x dy \; e^{y-x} \: |g(x)| \\
&=& \int_0^\infty dx \; (e^x-1) \: e^{-x} \: |g(x)| \;\leq\; 2 \:
	\|g \|_1
\end{eqnarray*}
shows that $f \in H^{1,1}(\R)$. Furthermore $dN_{|0} f = g$, and thus
$dN_{|0}$ is a bijection.
As consequence there are
neighborhoods $U \subset H^{1,1}(\R^+)$, $V \subset L^1(\R^+)$ of the
origin such that $N \::\: U \rightarrow V$ is a homeomorphism.

We now construct $\gamma$:
For $r_0$ large enough, $U(r)$ is positive and monotone for all
$r > r_0$. We introduce a new variable $s$ by
\[ s(r) \;=\; U^{-1} \left( \frac{r}{2} \right) \;\;\;\;,\spc
	s_0 \::=\: s(r_0) \]
and rescale \Ref{18},
\[ \gamma^\prime(s) \;=\; \sin(\gamma) + g(s) \;\;\;\;{\mbox{with}}\;\;\;\;
	g(s) \:=\: -\frac{\beta^\prime}{2U} \;\;\;\;,\spc
	s \geq s_0 \spc . \]
The function $g$ is in $L^1((s_0, \infty))$.
By further increasing $s_0$, we can arrange that
$g(\:\cdot\:-s_0) \in V$. Then $\gamma(r):=N^{-1}(g)(r+r_0) \in H^{1,1}((r_0,
\infty))$ satisfies \Ref{18} for $r \geq r_0$.
Finally we extend this solution to $H^{1,1}(\R^+)$.

To prove uniqueness let $\bar{\gamma}$ be a solution of \Ref{18} with
$\lim_{r \rightarrow \infty} \neq \pi \;({\mbox{mod }} 2 \pi)$.
Then $\bar{\gamma}(r) \in (-2C/r^2, \:2C/r^2) \;({\mbox{mod }} 2 \pi)$,
because otherwise $\bar{\gamma}$ will run into the attractive fixed point.
Thus $\bar{\gamma}(\:\cdot\:+r_0) \in V$ for $r_0$ large enough, and
the injectivity of $N_{|V}$ implies that $\bar{\gamma}=\gamma$.
\QED
We conclude that the equations \Ref{18}, \Ref{19} have (up to normalization)
a unique solution satisfying \Ref{20}.
We substitute \Ref{17a}, \Ref{17} into \Ref{2} and obtain for the ground 
state
the formula
\Equ{30}
\Psi^+(r, \varphi) \;=\; c \:\exp \left(-\int_0^r U(\tau) \:
	e^{-i \gamma(\tau)} \:d\tau \right) \left( \begin{array}{c}
	1 \\ e^{-i (\beta+\gamma)} \end{array} \right) \spc .
\EndEqu

\section{Discussion}
We come to a brief discussion of the ground state.
According to \Ref{30}, $\Psi^+$ is spherically symmetric, decays 
exponentially
at infinity and is regular at the origin.
We first look qualitatively at regions
where the eigenvectors of the superpotential do not depend much on $x$
(the eigenvalues can be arbitrary).
In this case $\beta^\prime$ is small and consequently,
$\gamma \approx 0$ (to see this, consider \Ref{18} backwards in $r$).
We thus have the formula
\[ \Psi^+(r, \varphi) \;\approx\; c \: \exp \left(
	-\int^r |W| \right) \: \left( \begin{array}{c}
	1 \\ e^{-i \beta} \end{array} \right) \spc . \]
$\Psi^+(r, \varphi)$ is proportional to the lowest eigenvector of the
superpotential, the factor $\exp(-\int |W|)$ is similar to the ground 
state of
$N=1$ quantum mechanics.
In regions where the eigenvectors of the superpotential are strongly
fluctuating, however, $\gamma$ can become large.
In this case, the spinorial dependence of $\Psi^+$ no longer minimizes
the expectation value of the superpotential.
The wave-function feels an ``effective'' potential described by the
additional phase factor $e^{-i \gamma}$ in \Ref{30}.
Nevertheless, the ground state energy remains zero.

We see that, in contrast to $N=1$ quantum mechanics, the existence of
a ground state does not depend on the degree of the superpotential.
On the other hand, our result is a confirmation for the hypothesis
expressed in \cite{JW} that the $N=1$ WZ-model on the cylinder has only
one ground state, even if the potential has several minima.

\end{document}